
\input phyzzx
\voffset = -0.4in
\footline={\ifnum\pageno=1 \nulline \else\newfootline \fi}
\def\nulline{{\hfill}}
\def\newfootline{\advance\pageno by -1\hss\tenrm\folio\hss}
\rightline{  UPR-647-T}
\rightline {February 1995}
\title {K\"{A}HLER POTENTIALS FOR ORBIFOLDS
 WITH CONTINUOUS WILSON LINES AND THE SYMMETRIES OF THE STRING ACTION}
\author{M. Cveti\v c$^{a }$, B. Ovrut$^{a }$ and \ W. A. Sabra
$^{b }$}
\address {$^{a}$
Department of Physics,\break
University of Pennsylvania,\break
Philadelphia,\break
PA 19104-6396,\break
U. S. A.\break}
\address {$^{b}$Department of Physics,\break
Royal Holloway and Bedford New College,\break
University of London,\break
Egham, Surrey, U. K.\break}
\abstract {By employing
the symmetries of the underlying conformal field theory,   the tree-level
K\"ahler potentials for untwisted moduli of the heterotic string
compactifications on orbifolds with continuous Wilson lines
are derived.
These symmetries act linearly on  bosonic (toroidal and $E_8\times E_8$ gauge)
string  coordinates  as well as on  the  untwisted  (toroidal and continuous
Wilson
lines) moduli; they correspond to the
scaling  of toroidal moduli, the   axionic shift of toroidal moduli  and  the
shift of the continuous Wilson line moduli.   In turn such symmetries provide
sufficient constraints to determine  the form of the low-energy effective
action
associated with the untwisted moduli up to a multiplicative factor.
 }
\endpage
\REF\one{L. Dixon, J. A. Harvey, C. Vafa and E. Witten, {\it Nucl. Phys.}
{\bf B261} (1985) 678; {\bf B274} (1986) 285.}
\REF\two{ A. Font, L. E. Ib\'a\~nez,
F. Quevedo and A. Sierra, {\it Nucl. Phys.} {\bf B331} (1991) 421.}
\REF\three{ R. Dijkgraaf, E. Verlinde and H. Verlinde, $On \ Moduli \ Spaces\
of\break
Conformal\  Field
\ Theories \ with\  c \geq 1$, Proceedings Copenhagen Conference,
Perspectives
in String Theory,
edited by P. Di Vecchia and J. L. Petersen,
World Scientific, Singapore, 1988.}
\REF\mir{ M. Cveti\v c, in Proceedings of ``Superstrings, Cosmology, and
Composite
Structures", College Park, Maryland, March 1987,
 S. J. Gates and R. Mohapatra ads.(World Scientific 1987); M. Cveti\v c.
{\it Phys Rev. Lett.} {\bf 59} (1987) 2829.}
\REF\four{ L. Dixon, D. Friedan, E. Martinec and S. H. Shenker,
{\it Nucl. Phys.} {\bf B282} (1987) 13;
S. Hamidi and C. Vafa,  {\em Nucl. Phys.} {\bf B279} (1987) 465.}
\REF\yukawa{D. Bailin, A.  Love and W. A.  Sabra, {\it Nucl. Phys.} {\bf B416}
(1994) 539.}
\REF\five{M. Cveti\v c, {\it Phys. Rev. Lett.}  {\bf 59}
(1987)  1795.}
\REF\thirteen{E. Witten, {\it Phys. Lett.} {\bf B155} (1985) 151;
S. Ferrara, C. Kounnas and M. Porrati, {\it Phys. Lett.} {\bf B181} (1986) 263;
S. Ferrara, L. Girardello, C. Kounnas and M. Porrati, {\it Phys. Lett.}
{\bf  B192} (1987) 368; {\it Phys. Lett.} {\bf B194}  (1987) 358. }
\REF\fourteen{H. P. Nilles. {\it Phys. Lett.} {\bf B180} (1986) 240;
C. P. Burgess, A. Font and F. Quevedo, {\it Nucl. Phys.} {\bf B272} (1986) 661;
U. Ellwanger and M. G. Shmidt, {\it Nucl. Phys.} {\bf B294 }(1987) 445; J.
Ellis, C. Gomez and D. V. Nanopoulos, {\it Phys. Lett.} {\bf B171}  (1986)
203.}
\REF\six{M. Cveti\v c, J. Louis and B. Ovrut, {\it Phys. Lett.} {\bf B206}, 229
(1998.}
\REF\seven{M. Cveti\v c, J. Molera and B. Ovrut, {\em Phys. Rev. D} {\bf 40}
(1989) 1140.}
\REF\fifteen{ S. Cecotti, S. Ferrara and L. Girardello, {\it Nucl. Phys.} {\bf
B308} (1989) 436;  {\it Phys. Lett.}  {\bf B213}  (1988) 443.}
\REF\threshold {G. L. Cardoso, D. L\"{u}st and B. A. Ovrut, hep-th-9410056,
 HUB-IEP-94/20, UPR-634T.}
\REF\ibanez{L. E. Ib\'{a}\~{n}ez, H. P. Nilles and F. Quevedo,
 {\it Phys. Lett.} {\bf B192} (1987) 332.}
\REF\mas{ L. E. Ib\'{a}\~{n}ez, J. Mas, H. P.  Nilles
and F. Quevedo, {\it Nucl. Phys.} {\bf B301} (1988) 157.}
\REF\eleven{L. Ib\'{a}\~{n}ez, W. Lerche, D. L\"{u}st and S. Theisen, {\it
Nucl. Phys.} {\bf B 352} (1991) 435.}
\REF\eight{G. L. Cardoso, D. L\"{u}st and T. Mohaupt,  {\it  Nucl. Phys.} {\bf
B432} (1994)
68.}
\REF\nine{T. Mohaupt, {\it  Int. J. Mod. Phys.} {\bf A9}  (1994)  4637.}
\REF\bailin{D. Bailin, A. Love, W. A. Sabra and S. Thomas,
 {\it QMW-TH}-94-28.}
\REF\ten{G. L. Cardoso, D. L\"{u}st and T. Mohaupt, HUB-IEP-94/50.}
\REF\sixteen{ K. S. Narain, {\it Phys. Lett.} {\bf  B169}  (1986) 41.}
 \REF\w{ K. S. Narain, M. H. Sarmadi and E. Witten,
         {\em Nucl. Phys.} {\bf B 279} (1987) 369.}
\REF\fer{S.  Ferrara,  P. Fre and P. Soriani,
 {\it Class. Quant. Grav.} {\bf  9} (1992)  1649.}
\REF\gilmore{R. Gilmore, Lie Groups, Lie Algebras and
some of Their Applications, \break Wiley--Interscience, New York, 1974.}
\REF\nineteen{S. Ferrara, C. Kounnas, D. L\"{u}st and
F. Zwirner, {\it Nucl. Phys.} {\bf B365} (1991) 431.}
\REF\seventeen{ M. Porrati and F. Zwirner, {\it Nucl. Phys.} {\bf B326} (1989)
162.}
\REF\eighteen{S. Ferrara and M. Porrati, {\it Phys. Lett. } {\bf B216} (1989)
216.}

The determination of a low-energy four-dimensional (4-d) lagrangian is a
necessary
step towards making phenomenological predictions from heterotic string theory,
such as the study of gauge coupling unification, the masses of  quarks
and leptons, supersymmetry breaking and string cosmology.
In particular,  heterotic string theories compactified on orbifolds  are of
phenomenological importance as they  give rise to an $N=1$ space-time
supersymmetric semi-realistic
4-d quantum field theories [\one, \two].
The orbifold models are characterized by a set of continuous parameters
describing the size and shape of the orbifold known
as toroidal moduli. They are the marginal deformations of the underlying
conformal field theory of the orbifold [\three] and they enter in
the space-time $N=1$  supersymmetric four dimensional Lagrangian
as chiral fields with  flat potentials to all orders in perturbation theory.
Toroidal moduli belong to  the untwisted sector of the orbifold. On the other
hand, the blowing-up modes, which parameterize the resolution of the orbifold
conical singularities belong to the twisted sector of the orbifold.

An appealing feature of orbifold compactification is that the toroidal moduli
can be included explicitly in the  calculations of the low-energy effective
action.\foot{In contrast, the explicit dependence on the blowing-up modes can
be
addressed only perturbatively in the value of  the blowing-up modes [\mir].}
thus allowing for probing the features of the effective
action in the whole sector of toroidal moduli.
The method to determine the explicit structure of the superpotential
of orbifold compactification directly from the underlying conformal field
theory was given in [\four]. The  explicit dependence
 on toroidal moduli has been calculated for
 Yukawa couplings\foot{See for example Ref. [\yukawa]  and references therein.}
and  non-renormalizable
terms [\five]  in the superpotential.
The explicit dependence of the  tree level  K\" ahler potential on the toroidal
moduli
has been evaluated in the literature
by several  methods. One method
relies on the truncation of the field theory limit of the string
[\thirteen, \fourteen]. Another method [\six, \seven] employs
the symmetries of the world-sheet action.
A similar method  has also been used in Ref.
[\fifteen].
In addition, significant progress has been made in addressing toroidal moduli
dependence
for genus-one (threshold)
 corrections\foot{See for example [\threshold]
and references therein.} to the low-energy action, in particular to the gauge
and gravitational couplings.

In  addition to the  toroidal moduli, the
untwisted sector of the  heterotic string theory compactified on orbifolds
also  contains Wilson line moduli, which arise when the
twist defining the orbifold is realized
 on the  $E_8\times E_8$ root  lattice by a rotation [\ibanez, \mas].
The inclusion of Wilson line moduli is essential as
they lead  to models with semi-realistic gauge groups (of lower rank)
via the \lq\lq stringy Higgs effect" [\eleven], and thus  may provide a class
of orbifold compactifications with a semi-realistic features.

Recently, there has been a renewed interest in addressing the low-energy
effective lagrangian  for  orbifolds with continuous Wilson lines
[\eight, \nine, \bailin, \ten].  As a first step one would like to address
the explicit dependence of the K\" ahler potential on the
Wilson line moduli, thus allowing for the full description of the untwisted
sector of the orbifold moduli space.   In  Ref. [\eight]
the local structure of the moduli space of ${\bf Z}_N$ orbifolds with
continuous Wilson lines was obtained, by using the
compatibility between the Narain twist and the toroidal
moduli, {\it i.e.}, one starts with the moduli space of toroidal
compactification [\sixteen] and then determine which subspace
is compatible with the action of the orbifold twist
 on the underlying Narain lattice.

The  purpose of  this letter is to derive the explicit
 dependence of the  K\"{a}hler potentials on the full
untwisted sector of the  orbifolds with
 continuous Wilson lines  by using the symmetries of the
the underlying conformal field theory which is
 associated with the full untwisted moduli sector of
the corresponding orbifold compactification.
This method  provides a generalization of the one in [\six ,\seven] which
only addresses  the  K\"{a}hler potential for
toroidal moduli and untwisted matter fields.

We start our discussion with the heterotic string theory
compactified on a torus. A $d$-dimensional torus ($d=6$)
can be defined as a quotient of ${\bf R}^d$ with respect to a
lattice $\Lambda$ defined by
$$\Lambda=\{\sum_{i=1}^d a^i \hbox {e}_i,\ \ \   a^i \in Z\},
\qquad i=1,\cdots d\ \ .\eqn\hell$$
The world-sheet action associated with the bosonic string
coordinates is of the form [\w]:
$${\cal S}={1\over{2\pi}}\int dz d \bar z \Big(b_{ij}\partial\phi^j
\bar \partial\phi^i+A_{Ij}\partial\phi ^j \bar\partial\varphi^I +C_{IJ}
\partial\varphi^J \bar\partial\varphi^I\Big)\ \ ,\eqn\action$$
supplemented with the constraint
$C_{IJ}\partial\varphi^J+A_{Ij}\partial\phi^j=0.$ Here,
 $z$ ($\bar z$) and $\partial$ ($\bar \partial$) correspond to the left
(right)-moving world-sheet
coordinates and the corresponding partial derivative, respectively, $b_{ij}$ is
the background metric denoting
the metric and the antisymmetric tensor coefficients (toroidal moduli),
$A_{Ij}$ is the Wilson line and $C_{IJ}$ is the Cartan metric of $E_8\times
E_8$.
Denote the internal toroidal string  coordinates  by
the $d$-dimensional column matrix $\phi$ and the  $E_8\times E_8$ bosonic
string coordinates  by the $D$-dimensional  ($D=16$)
column matrix by $\varphi.$
The equations of motion plus the chiral constraint can be written
 in  matrix notation in the following form:
$$\bar\partial G=0,\qquad {G}=bF+A^tC{\cal F}\eqn\cl$$
$${\cal G}=C{\cal F}+CAF=0\ \ ,\eqn\con$$
where  $F=\{\partial \phi^j\}$, ${\cal F}=\{\partial\varphi^J\}$, $b$ is a
$d\times d$
matrix representing the metric and  antisymmetric tensor
background fields $b_{ij}$, $A$ is $D\times d$ matrix representing the
Wilson line moduli $A_{\ j}^I$ , and $C$ is the Cartan matrix  ($C_{IJ}$) of
$E_8\times E_8$.

The equations of motion have the following symmetries.
First,  there is  a rescaling symmetry given as
$$\eqalign{ b&\rightarrow {(L^{-1})}^t\, b\, L^{-1},\cr
 A&\rightarrow AL^{-1},\cr F&\rightarrow LF,\cr {\cal F}&\to{\cal F} ,}
   \eqn\vv$$
 where  $L$ is an  invertible constant $n\times n$ matrix.

Second, there is the  well known symmetry under the
axionic shift:
$$\eqalign{b&\rightarrow b+N \ \ ,\cr A &\rightarrow A \ \ , \cr  F&\rightarrow
F\ \ ,\cr
{\cal  F}&\rightarrow {\cal F},}\ \ \eqn\vvv$$
where $N=-N^t.$
The symmetries \vv\ and \vvv\ are those discussed in Ref. [\six]
and  are used to fix the form of the kinetic energy term of the $b$ moduli
up to an overall constant,  which is   determined  to be 1 (in Planck units)
by   calculating the   string scattering amplitude  with  four toroidal moduli
vertex
insertions. Consequently,  the form of the
corresponding K\" ahler potential is fully  determined.

In our case, an additional symmetry, which  relates the toroidal
and Wilson line moduli, is needed. It  is obtained  by imposing
a constant shift $W$ on the Wilson line moduli $A$, along with  the following
transformation
on the toroidal moduli $b$ as well as on the bosonic string coordinates:
$$\eqalign{A&\rightarrow A+W\ \ ,\cr b&\rightarrow b+W^t{C}A+A^t
CW+W^t{C}W\ \ ,\cr
F&\to F\ \ ,\cr
 {\cal F}&\rightarrow{\cal F}-W F\ \
.}\eqn\pinhead$$
The constraint equation \con\ is crucial to prove that the
equation of motion \cl\ is invariant under the symmetry  transformations
\pinhead.\foot{Note,  the symmetry transfomations \vv -\pinhead\
are linear in the background fields.}

The above sets of symmetry transformations
\vv-\pinhead\ correspond to the symmetries of the underlying conformal field
theory
and therefore correspond to the symmetry for the
corresponding  low-energy effective action associated with the  untwisted
moduli sector. These symmetries are sufficient to
 determine the tree-level
kinetic terms of the untwisted moduli sector up to an overall constant.
 Namely, the kinetic energy of
of the untwisted sector moduli is of the form:
$${\cal K}={{\partial^2  K}\over{\partial T_i\partial T_j}}
\partial_\mu T_i\partial^\mu T_j\ \ ,\eqn\kin$$
where $T_i$ correspond to the  untwisted moduli fields in $b$ and $A$,
where $K$  is,
up to an multiplicative  constant, of the following form:
$$ K = -{\rm log \ det} (b+b^t-2A^tCA)\ \ .\eqn\k$$
The multiplicative constant can be determined by a calculation of the
string amplitude  with four toroidal moduli vertex insertions.

The  strucutre of the  coset space associated with the  metric
${{\partial^2  K}\over{\partial T_i\partial T_j}}$  (defined in \kin -\k )
is of the type $SO(d,d+D)/SO(d)\otimes SO(d+D)$ ($d=6, \ D=16$).

We would also  like to point out  that the
equations of motion {\cl }-{\con}  actually possess
a larger symmetry [\fifteen] than the one specified by symmetry
transformations  \vv-{\pinhead}.
The larger symmetry   corresponds in  general to  a subgroup of
$SO(d+D, d+D)$ ($d=6, \ D=16$) which is consistent with the constraints ${\cal
G}=0$ (Eq. {\con}).
If one  represents an element of $SO(d+D,d+D)$ with
$$g=\pmatrix{g_1&g_2\cr g_3&g_4}\ ,\qquad g^t\Omega g=\Omega\ ,
\qquad \Omega=\pmatrix{0&1\cr 1&0} , \eqn\dd$$
then the group  element has the following
action on the  moduli  (background) fields and  the
 corresponding bosonic string coordinates:
$$\pmatrix{F\cr {\cal F}\cr G\cr
 {\cal G}}\rightarrow g\pmatrix{F\cr {\cal F}\cr G\cr {\cal G}}\ \ .\eqn\st$$
The symmetry transformations {\vv}-{\pinhead}   correspond to  a subset of
those specified in  Eq. {\st},  and thus, can be obtained by setting:
$$ g_1=\pmatrix{\alpha_1&0\cr \alpha_2&1},\qquad g_2=0,\qquad
g_3=\pmatrix {\gamma_1&0\cr 0&0}\ ,\quad \hbox{with}\  \gamma_1=-\gamma_1^t
\ .\eqn\hia$$
Using \dd\ and \hia\ we obtain the following explicit form of the subset of the
group elements:

$$g=\pmatrix{\alpha_1&0&0&0\cr\alpha_2&1&0&0\cr\gamma_1&0&
(\alpha_1^{-1})^{t}&-{({\bf }\alpha_1^{-1})}^{t}\alpha_2^t\cr 0&0&0&1}\
,\qquad\eqn\gg$$
which  in turn  transforms for the  untwisted moduli in the following
way $$\eqalign{b&\rightarrow \gamma_1\alpha_1^{-1}+(\alpha_1^{-1})^t
b\alpha_1^{-1}-
{(\alpha_1^{-1})}^t{(\alpha_2)}^t CA\alpha_1^{-1}-{(\alpha_1^{-1})}^tA^t
C\alpha_2\alpha_1^{-1}\cr &
+{(\alpha_1^{-1})}^t\alpha_2^tC\alpha_2\alpha_1^{-1} \ ,\cr
A&\rightarrow A\alpha_1^{-1}-\alpha_2\alpha^{-1}_1 \ \ .}\eqn\tran$$
Obviously, these transformations constitute the rescaling \vv , axionic shift
\vvv , and the shift of the Wilson lines \pinhead .

The analysis for the toroidal  case can now be generalized in a straightforward
way to the case of orbifolds with continuous Wilson lines.
A symmetric $d$-dimensional($d=6$) ${\bf Z}_N$ orbifold [\one] can be
obtained by identifying points on the ${\bf T}^d$ torus
under the action of a cyclic group ${\bf Z}_N$, generated by the twist
$\theta$,
$${\bf Z}_N=\{\theta^j, j=1,...,N-1\}\ \ .\eqn\tw$$
In order to allow for the presence of continuous Wilson line moduli,
the twist should also act as a rotation on the $E_8\times E_8$ lattice
coordinates.
Let the action of the twist on the toroidal string  coordinates be represented
by
 ($d\times d$) matrix $Q$ and  on the gauge coordinates by the ($D'\times
D'$)  matrix   $M$, with  $D'$ the dimension of  subspace of  the $E_8\times
E_8$ lattice on which the twist has non-trivial action.  Then for
the untwisted moduli  to  be consistent with the action of the twist,
they have to  satisfy the following conditions [\nine, \bailin]:
$$Q^tbQ=b\ , \qquad MA=AQ\ \ .\eqn\c$$
Due to the above conditions, the orbifold has fewer moduli than the
corresponding torus and thus the string action  for the untwisted sector of
the ${\bf Z}_N$ orbifolds contains  the
subset of terms in the toroidal string
action \action  (supplemented by the chiral constraint \con ),
which  are invariant under the corresponding twist transformations.
Thus, the symmetry transformations
of the underlying conformal field theory
constitute a  subset of transformations \vv -\pinhead  ,
now acting on the remaining moduli and the string coordinates on which the
twist has a non-trivial action.
Consequently, the  effective  4-dimensional kinetic terms of the moduli fields
and the corresponding
K\" ahler potential
of  all ${\bf Z}_N$ orbifolds can be determined  explicitly.
The  coset spaces spanned  by  the untwisted  moduli of ${\bf Z}_N$ orbifolds
correspond to K\" ahler spaces, which are  subspaces of
$\textstyle {SO(d,d+D)}\over \textstyle{{SO(d)\otimes SO(d+D)}}$ ($d=6,\
D=16$).

For the discussion of orbifolds  it is in certain cases  more convenient to
represent  the
$d$-toroidal ($\phi^j$) and  $D$-gauge ($\varphi^I$)
real string coordinates in a complex basis by $Z^a$ and ${\cal Z}^{m}$  string
coordinates, respectively. A twist of order $N$ acts on the complex coordinates
as a multiplication
by $e^{2\pi i/ N}$.
The complex coordinates
are related to the real ones via the relation [\fer]:
$$\eqalign{\phi^i=&Z^a E_a^i+{\bar Z}^{a*}E_{a^*}^i\ ,\qquad a, a^*=1,2,. . . ,
{d\over 2 }(=3)\ \ ,\cr
\varphi^I=&{\cal Z}^m {\cal E}_m^I+{\bar {\cal Z}}^{m*}{\cal E}_{m^*}^I\
,\qquad
m, m^*=1,2 \cdots,  {D'\over 2}\ \ ,}\ \ \ \eqn\com$$
where $E_a$, $E_{a^*}$ are the complex basis vectors
of the orbifold and
${\cal E}_m$, ${\cal E}_{m^*}$ are those of a $D'$-dimensional ($D'\le D=16$)
 subspace of  the $E_8\times E_8$ lattice.
Using the same symmetries  \vv -\pinhead\ as before, however,
 now all expressed in the complex basis,
the K\" ahler potential can be written as
$$K=- {\rm log\  det} ({\bf b}+{\bf b}^\dagger-2{\bf A}^\dagger{\bf C}{\bf
A})\ \ ,\eqn\ll$$
where  $\bf b$ and $\bf A$ are complex matrices representing the background
metric and Wilson lines and $\bf C$ is the Cartan metric expressed in
the complex basis.

Thus, the corresponding subsets of symmetry transformations \vv -\pinhead\
provide us with  the determination  of the  effective  4-dimensional  action,
{\it i.e.}, the kinetic energy, the
  local  structure of the  coset spaces spanned by the toroidal and
 continuous Wilson line moduli of ${\bf Z}_N$ - orbifolds.
In particular,  the  untwisted subsector of  ${\bf Z}_N$ -orbifolds  associated
with
a   two-torus (${\bf T}^2$) modded out  by  a $Z_2$ twist,
 has the {\it same} moduli  as the  corresponding two-torus, {\it i.e.}, there
are  torodial moduli (four real  fields) $b_{ij}$ ( $(i,j)=1,2$) and Wilson
line
moduli whose number depends on the choice of the gauge twist.
The  corresponding kinetic energy for these moduli
is of the type \kin-\k.
The coset space  spanned by the moduli is the K\"ahler space [\gilmore]  of the
type
$\textstyle{SO(2,r)}\over\textstyle{{SO(2)\otimes SO(r)}}$.
The  set of complex coordinates for these
 cosets can be constructed  [\seventeen, \eighteen, \nineteen, \eight]
 in terms of the real  toroidal and Wilson line moduli.
The explict transformation
between the four real toroidal moduli and four Wilson line moduli, and
the  corresponding  two complex toroidal and two complex Wilson line moduli,
representing the coset $\textstyle{SO(2,4)}\over\textstyle{{SO(2)\otimes
SO(4)}}$,
was given in Ref. [\eight].\foot{ The space with the  toroidal
moduli $b_{ij}$, only,  corresponds  the coset space
${SO(2,2)\over{SO(2)\otimes SO(2)}}\equiv \left[{SU(1,1)}\over
{U(1)}\right]^2$.
The explict transformation between the real fields $b_{ij}$
and the  corresponding  two-complex  toroidal moduli
was given in  Ref [\six]. For related transformations with inclusion of
untwisted
matter fields see Refs. [\seventeen, \eighteen, \nineteen].}

In order to further illustrate the method, we  consider a $2$-dimensional
($d=2$) ${\bf Z}_3$ orbifold constructed from the torus
$\textstyle{R^2}\over\textstyle{\Lambda}$, where $\Lambda$ is the root lattice
of $SU(3)$,
and for the sake of simplicity we take the
gauge twist to act on an $SU(3)$ subgroup of the $E_8\times E_8$.
 The action of the twists on the toroidal and the  subset of the
$E_8\times E_8$ bosonic coordinates are represented by:
$$Q=\pmatrix{0&-1\cr1&-1}\ ,\quad M=\pmatrix{0&-1\cr1&-1} \ \ .\eqn\eva$$
Because of condition \c\  this orbifold has one independent Wilson line modulus
[\eight]. The  Cartan matrix  $C_{IJ}$ of $SU(3)$ and  Wilson line
 fields  $A_{\ j}^I$ are represented in the real lattice basis
by the following matrices:
$$C=\pmatrix{2&-1\cr -1&2}\ ,\quad
 A=\pmatrix{A_{\ 1}^1&A_{\ 2}^1\cr A_{\ 1}^2&A_{\  2}^2}\ \  .\eqn\realmat$$
Recall, the lower  and upper indices  of $A_{\ j}^I$ refer to the toroidal
(target
space) and  gauge  indices, respectively.
Using the consistency conditions \c\  with the twists represented as \eva,
the background metric $b$  and $A$ can be represented as
$$A=\pmatrix{p&q\cr -q&p+q}\ ,\quad
b=\pmatrix{g_{11}&-g_{11}/2-B\cr -g_{11}/2+B&g_{11}}\ \ .\eqn\sophie$$
The complex basis of the orbifold and the gauge coordinates are given by
$$\eqalign{E_1=&{\hbox {e}}_1-\alpha{\hbox {e}}_2\ ,\cr
{\cal E}_1=&e_1-\alpha e_2 \ ,\qquad  \alpha=e^{2i\pi/3}}\ \ .\eqn\complex$$
 The complex coordinates are given in terms of the real coordinates by
$$\eqalign{Z^1=&{\alpha^2\over(\alpha^2-\alpha)}\phi^1+
{1\over(\alpha^2-\alpha)}\phi^2\ \ ,\cr
{\cal Z}^1=&{\alpha^2\over(\alpha^2-\alpha)}\varphi^1+
{1\over(\alpha^2-\alpha)}\varphi^2 \ \ .}\eqn\london$$
The world-sheet action (for the twisted coordiates) can now be written as
$${\cal S}={1\over {2\pi }} \int dz d \bar z \Big({\bf b}\partial Z^1
\bar\partial{\bar Z}^{1^*}+{\bf A}\partial Z^1
\bar\partial\bar{\cal Z}^{1^*} +{\bf C}
\partial{\cal Z}^{1}\bar\partial\bar{\cal Z}^{1^*}\Big)+ h.c.\ \ .\eqn\susan$$
Notice that this is the most general world-sheet action consistent with the
action of both the toroidal and gauge twists.
The complex moduli are expressed in terms of the real ones by
$$\eqalign{& {\bf b}=
{3\over 2}g_{11}+i{\sqrt 3} B=\sqrt{ 3}(\sqrt{\det g}+iB)\equiv {\sqrt 3}T\
,\cr
&{\bf A}=3(p-\alpha q)\equiv {3 \over {\sqrt 2}} {\cal A}\  , \cr
& {\bf C}=3
\ \ .}\eqn\complex$$

Using \ll, the K\" ahler potential can be written, up to an overall positive
multiplicative factor, in the  following complex form :
$$K= -{\rm log }(T+ T^*-{\sqrt 3}{\cal A}^*{\cal A})+ const.\ \
,\eqn\hampsted$$
This is the K\" ahler potential for the
 $\textstyle{SU(1,2)}\over \textstyle{SU(2)\otimes U(1)}$ coset space.

If, for example,  one allows the gauge twist to act on another $SU(3)$ subgroup
 with the eigenvalue $e^{4\pi i/2}$, {\it i.e.},  in the real basis the twist
is generated by $M'=\pmatrix{-1&1\cr -1&0}$,
 then the moduli space will contain
 another  complex Wilson line modulus.
Denote ${\cal E}_2$ the complex vector representing  the second $SU(3).$
In the real basis the Wilson line is  given by
$$A=\pmatrix{p&q\cr -q&p+q\cr p'&q'\cr p'+q'&-p'}.\eqn\york$$

The world-sheet action in this case can be written as
 $$\eqalign{{\cal S}={1\over{2\pi}} \int dz d \bar z &\Big({\bf b}\partial Z^1
\bar\partial{\bar Z}^{1^*}+{\bf A}\partial Z^1
\bar\partial\bar{\cal Z}^{1^*} +{\bf C}
\partial{\cal Z}^{1}\partial\bar{\cal Z}^{1^*}\cr
&+{\bf C'}
\partial{\cal Z}^{2}\bar\partial\bar{\cal Z}^{2}+{\bf A'}\partial Z^1
\bar\partial{\cal Z}^{2} \Big)+ h.c. \ ,} \eqn\sus$$
where ${\bf A'}=3(p'-\alpha q')\equiv {3\over{\sqrt 2}}{\cal A}^\prime$ and
${\bf C'}= 3$.
It is then straightforward to show that  the K\" ahler potential can be written
as
$$K=-{\rm  log} (T+ T^*-{\sqrt 3}{\cal A}^*{\cal A}-{\sqrt 3}{\cal
A}^{\prime *}{\cal A}^{\prime})\ \ .\eqn\f$$
This is the K\"ahler potential for the $\textstyle{SU(1,3)}\over
 \textstyle{SU(3)\otimes U(1)}$ coset space.

In general,
$\textstyle{SU(1,k)}\over \textstyle{SU(k)\otimes U(1)}$
coset space is parametrized by one complex modulus $T$ and $k-1$ complex
Wilson line moduli. This  coset space corresponds to the  untwisted moduli
sector
of any  2-dimensional ${\bf Z}_N (N\not=2$) orbifold. Therefore,
it also corresponds to a subsector of $d$-dimensional ($d=6$) ${\bf Z}_N$
orbifold, which  corresponds to the untwisted moduli sector of
a two-torus modded out by a $Z_{N}$ twist represented by a
 complex  phase. In  models  corresponding to the  string vacua
the number of the complex Wilson lines
 of course depends on the choice of the gauge twist which is constrained
 by  the world-sheet modular invariance [\ibanez, \mas].
Applying the  method presented  to all ${\bf Z}_N$ orbifolds, one obtains
the same coset spaces as those in [\seven, \eight]
with the Wilson lines playing the
role  similar to those of  untwisted matter fields.

In conclusion, we have presented a  method to determine the K\"{a}hler
potential  of the moduli fields, {\it i.e.}, toroidal moduli as well as
continuous Wilson lines, in the untwisted sector of the
heterotic string theory compactified on  any ${\bf Z}_N$ orbifold.
The method employs the  symmetries of the underlying conformal field
theory  associated with the  untwisted moduli sector of the theory.
These symmetries act linearly on  bosonic (toroidal and $E_8\times E_8$ gauge)
string  coordinates  as well as the  untwisted  (toroidal and continuous Wilson
lines) moduli; they correspond to the
scaling of toroidal moduli, the axionic shift of toroidal moduli and  the
shift of the continuous Wilson line moduli. Such symmetries provide  sufficient
constraints to determine  the form of the low-energy effective action
 of the  untwisted moduli sector up to a multiplicative factor.
In turn,  the local structure of the
 K\" ahler manifold associated with of   untwisted moduli sector   can be fully
determined.

{\bf Acknowledgments}
We would like to thank   S. Thomas for useful discussions.
M. C. and W. S would  also like to thank organizers of the 1994 U.K. Institute
for
Theoretical Physics, where the work was initiated, for hospitality.
The work is  supported by  U.S. DOE
Grant No. DOE-EY-76-02-3071,  the NATO collaborative research grant CGR
940870 (M.C., B. O)  and PPARC (W.S.).
\refout
\end